\documentclass[11pt]{article}

\usepackage{latexsym,amsmath,amssymb,amscd,eucal}
\usepackage{xypic}
\usepackage{enumerate}

\def\harr#1#2{\smash{\mathop{\hbox to .3in{\rightarrowfill}}
 \limits^{\scriptstyle#1}_{\scriptstyle#2}}}

\def\appendix#1{\addtocounter{section}{1}\setcounter{equation}{0}
\renewcommand{\thesection}{\Alph{section}}
\section*{Appendix \thesection\protect\indent \parbox[t]{11.715cm} {#1}}
\addcontentsline{toc}{section}{Appendix \thesection\ \ \ #1} }
\newcommand{\eq}{\begin{equation}}
\newcommand{\eqend}{\end{equation}}

\newbox\ncintdbox \newbox\ncinttbox
\setbox0=\hbox{$-$} \setbox2=\hbox{$\displaystyle\int$}
\setbox\ncintdbox=\hbox{\rlap{\hbox to
\wd2{\hskip-.125em\box2\relax \hfil}}\box0\kern.1em}
\setbox0=\hbox{$\vcenter{\hrule width 4pt}$}
\setbox2=\hbox{$\textstyle\int$}
\setbox\ncinttbox=\hbox{\rlap{\hbox to
\wd2{\hskip-.175em\box2\relax \hfil}}\box0\kern.1em}

\def\be{\begin{equation}}
\def\ee{\end{equation}}
\def\bea{\begin{eqnarray}}
\def\eea{\end{eqnarray}}
\def\bd{\begin{displaymath}}
\def\ed{\end{displaymath}}

\DeclareFontFamily{U}{rsf}{}
\DeclareFontShape{U}{rsf}{m}{n}{
  <5> <6> rsfs5 <7> <8> <9> rsfs7 <10-> rsfs10}{}
\DeclareMathAlphabet\Scr{U}{rsf}{m}{n}

\makeatletter
\newdimen\normalarrayskip              
\newdimen\minarrayskip                 
\normalarrayskip\baselineskip
\minarrayskip\jot
\newif\ifold             \oldtrue            
\def\arraymode{\ifold\relax\else\displaystyle\fi} 
\def\@arrayskip{\ifold\baselineskip\z@\lineskip\z@
     \else
     \baselineskip\minarrayskip\lineskip2\minarrayskip\fi}
\def\@arrayclassz{\ifcase \@lastchclass \@acolampacol \or
\@ampacol \or \or \or \@addamp \or
   \@acolampacol \or \@firstampfalse \@acol \fi
\edef\@preamble{\@preamble
  \ifcase \@chnum
     \hfil$\relax\arraymode\@sharp$\hfil
     \or $\relax\arraymode\@sharp$\hfil
     \or \hfil$\relax\arraymode\@sharp$\fi}}
\def\@array[#1]#2{\setbox\@arstrutbox=\hbox{\vrule
     height\arraystretch \ht\strutbox
     depth\arraystretch \dp\strutbox
     width\z@}\@mkpream{#2}\edef\@preamble{\halign \noexpand\@halignto
\bgroup \tabskip\z@ \@arstrut \@preamble \tabskip\z@ \cr}%
\let\@startpbox\@@startpbox \let\@endpbox\@@endpbox
  \if #1t\vtop \else \if#1b\vbox \else \vcenter \fi\fi
  \bgroup \let\par\relax
  \let\@sharp##\let\protect\relax
  \@arrayskip\@preamble}
\makeatother

\newcommand{\beq}{\begin{eqnarray}}
\newcommand{\eeq}{\end{eqnarray}}

\def\appendix#1{\addtocounter{section}{1}\setcounter{equation}{0}
\renewcommand{\thesection}{\Alph{section}}
\section*{Appendix \thesection. #1}
\addcontentsline{toc}{section}{Appendix \thesection\ \ \ #1} }

\newtheorem{theorem}{Theorem}[section]

\newtheorem{notation}{Notation}[section]

\newtheorem{definition}{Definition}[section]

\newtheorem{remark}{Remark}[section]

\numberwithin{equation}{section}

\begin{document}


\vspace{.1in}

\begin{center}

{\Large\bf TRUNCATED HEAT KERNEL AND ONE-LOOP DETERMINANTS FOR THE
BTZ GEOMETRY}

\end{center}
\vspace{0.1in}
\begin{center}
{\large
A. A. Bytsenko $^{(a)}$
\footnote{abyts@uel.br}}
and M. E. X. Guimar\~aes $^{(b)}$ \footnote{emilia@if.uff.br}
\vspace{7mm}
\\
$^{(a)}$ {\it
Departamento de F\'{\i}sica, Universidade Estadual de
Londrina\\
Caixa Postal 6001, Londrina-Paran\'a, Brazil}
\vspace{5mm}\\
$^{(b)}$ {\it Instituto de F\'{\i}sica,
Universidade Federal Fluminense,\\
Av. Gal. Milton Tavares de Souza s/n, Niter\'oi-RJ, Brazil}\\
\end{center}
\vspace{0.1in}
\begin{center}
{\bf Abstract}
\end{center}
There is a special relation between the spectrum and the {\it truncated} heat kernel of the Euclidean BTZ black hole with the Patterson-Selberg zeta function. Using an orbifold description of this relation we calculate the on-shell corrections of the gravitational quantum fluctuations.

\vfill

{Keywords: .........}


\newpage
\tableofcontents


\section{Introduction}

One of the most important issue in studying quantum gravity is the
black hole physics, the origin of the Bekenstein-Hawking entropy
and its quantum corrections. Recently quantum gravity partition
functions in three dimensions have been analyzed in detail
\cite{Maloney,Giombi}. The three dimensional case is quite simple
(because of no propagating gravitons) but there is a common belief
that it deserves attention  as an useful laboratory as analog
models for four dimensional physics. A simple geometrical
structure of three-dimensional black hole (Ba\~nados, Teitelboim, Zanelli (BTZ) black hole \cite{Banados}) allows
exact computations since its Euclidean counterpart is locally
isomorphic to the constant curvature hyperbolic space $H^3$.

In physical literature usually assumes that the fundamental domain for the action of a discrete group $\Gamma$ has finite volume. On the other hand BTZ black hole has a Euclidean quotient representation ${\bf H}_{\Gamma} = \Gamma\backslash H^3$ for an appropriate $\Gamma$, where the fundamental domain has {\it infinite} hyperbolic volume.
\footnote{
For the non-spining black hole one can choose $\Gamma$ to be Abelian group generated by a single hyperbolic element
\cite{Perry}.} 
For discrete groups of isometries of three-hyperbolic space with infinite volume of fundamental domain (i.e. for Kleinian groups)
Selberg zeta functions and trace formulas, excluding fundamental domains with cusps, have been considered in \cite{Perry1}, where
results depends on previous works \cite{Patterson1}, \cite{Patterson2}. Note that matters are difficult in case of infinite-volume setting due to the infinite multiplicity of the continuous spectrum and absence of a canonical renormalization of the scattering operator which makes it trace-class. However, for BTZ black hole one can by-pass much of the general theory and proceed more directly to define a Selberg zeta function attached to ${\bf H}_ {\Gamma}$ and establish a trace formula which is a special version of the Poisson formula for resonans (see for detail \cite{Perry}).

In \cite{Bytsenko} the one-loop correction to the
Bekenstein-Hawking entropy for the non-spinning black hole were studied by
expressing the determinants in terms of the appropriate heat
kernels, which were evaluated using a method of images (for the
analogous derivation of the Selberg trace formula see 
\cite{Bytsenko2}). However, this correction to the entropy is not completely corrected. Indeed, the procedure of regularization
of the divergent volume of the fundamental domain, which has been
made in \cite{Bytsenko}, changes the group actions on the real
hyperbolic three-space, and cannot be compatible with  the origin
structure of the cyclic groups. In fact, there is a special
relation between the spectrum and the {\it truncated} heat kernel
of the Euclidean BTZ black hole with the Patterson-Selberg zeta
function \cite{Bytsenko1}. The main purpose of this work is to
provide the corrected relation between the spectral functions and the truncated heat kernel for the BTZ black hole, and calculate the one-loop quantum correction to the partition function.

\section{Three-dimensional black hole}

The Euclidean three-dimensional black hole has an orbifold
description ${\bf H}_{\Gamma_{(a,b)}}=\Gamma_{(a,b)}\backslash
H^{3}$ if we choose the suitable parameters $a>0$, $b\geq 0$,
where $H^{3}=\{(x,y,z)\in\mathbb{R}^3\mid z>0\}$ is an hyperbolic
three-dimensional space (or simply a three-space) and
$\Gamma_{(a,b)}\subset SL(2,\mathbb{C})$ is a cyclic group of
isometries.  Space ${\bf H}_{\Gamma_{(a,b)}}$ is a solution of the
Einstein equations $ R_{ij}- \frac{1}{2}g_{ij}R_{g}-\Lambda
g_{ij}=0 $ with hyperbolic metric $ ds^{2}=
(\sigma/z)^{2}(dx^2+dy^2+dz^2) $ and negative cosmological
constant $\Lambda$, $\sigma=(-\Lambda)^{-\frac{1}{2}}$, and
constant scalar curvature $R_{g}= 6\sigma^{-2}=-6\Lambda$. The
original BTZ metric in the coordinates $(r,\phi, t)$ can indeed
be transformed to the metric $ds^2$ by a specific change of
variables $(r, \phi, t)\rightarrow (x,y,z)$. It is the periodicity in the Schwarzschild
variable $\phi$ that allows for the above orbifold description. In
fact, the parameters $a, b$ are given as follows. For $M>0$,
$J\geq 0$ the black hole mass and angular momentum, and for
$r_{+}>0, r_{-}\in i\mathbb{R}$ ($i^2=-1$) the outer and inner
horizons given by
\begin{equation}
r_{+}^2=\frac{M\sigma^2}{2}\left[1+\left(1+\frac{J^2}{M^2\sigma^2}\right)^{\frac{1}{2}}\right]\,,\,\,\,\,\,\,\,\,\,\,
r_{-} = -\frac{\sigma Ji}{2r_{+}}\,.
\label{eq: rs}
\end{equation}
One obtains
$
a:=\pi r_{+}/\sigma,
$
$
b:=\pi\left|r_{-}\right|/\sigma.
$
\begin{definition}
$\Gamma_{(a,b)}$ is defined to be the cyclic
subgroup of $G=SL(2,\mathbb{C})$ with generator
\begin{eqnarray}\label{eq: gammaGammaab}
\gamma_{(a,b)}&\stackrel{{ def}}{=}& \left[
\begin{array}{cc}e^{a+ib} & 0 \\ 0 &
e^{-(a+ib)}\end{array}\right]\,,
\label{g1}
\\
\Gamma_{(a,b)} &\stackrel{{ def}}{=}& \{\gamma^n_{(a,b)}\mid n\in\mathbb{Z}\}\,.
\label{G1}
\end{eqnarray}
\end{definition}
The Riemannian volume element $dV$ corresponding to the metric
$ds^2$ is given by $ dV=\frac{\sigma^3}{z^3}dxdydz. $ It is known
that a fundamental domain ${F}_{(a,b)}$ for the action of
$\Gamma_{(a.b)}$ on $H^3$ is given by $ F_{(a,b)}=\{(x,y,z)\in H^3
\mid 1<x^2+y^2+z^2<e^{2a}\}. $ It follows that $\Gamma_{(a,b)}$ is
a Kleinian subgroup of $G$ and
\begin{equation}
{\rm Vol}\left(F_{(a,b)}\right)=\int_{F_{(a,b)}}dV =\infty.
\end{equation}

{\bf Scattering matrix and resonances.}
Since $F_{(a,b)}$ has an infinite hyperbolic volume, the usual
spectral theory for the Laplacian $\Delta_{\Gamma_{(a,b)}}$ of
${\bf H}_{\Gamma_{(a,b)}}$ does not apply (as it does for finite
volume orbifolds).  We outline, briefly, a suitable spectral
analysis of $-\Delta_{\Gamma_{(a,b)}}$ where a key notion is that
of scattering resonances (see for more detail \cite{Perry}).
Henceforth we shall write $\Gamma$ for $\Gamma_{(a,b)}$;
one notes that $\Delta_\Gamma$ is given by
\begin{equation}
\Delta_\Gamma=
\frac{1}{\sigma^2}\left[z^2\left(\frac{\partial^2}{\partial
x^2}+\frac{\partial^2}{\partial y^2} +\frac{\partial^2}{\partial
z^2}\right)-z\frac{\partial}{\partial z}\right]\,.
\end{equation}
The space of square-integrable functions on the black hole
${\bf H}_\Gamma=\Gamma\backslash H^3$, with respect to the measure $dV$, has a orthogonal decomposition
\begin{equation}
L^2\left({\bf H}_\Gamma,dV\right)=\sum_{m,n\in\mathbb{Z}}\oplus H_{mn} \,\,\,\,\,
{\rm with}\,\,\,\,\, {\rm Hilbert}\,\,\,\,\, {\rm space}
\,\,\,\,\, {\rm isomorphisms}\,\,\,\,\, H_{mn}\simeq
L^2\left(\mathbb{R}_{+},dt\right)\,,
\end{equation}
for $\mathbb{R}_{+}$ the space of positive real numbers.
A spectral decomposition becomes
\begin{equation}
-\sigma^2\Delta_\Gamma\simeq\sum_{m,n\in\mathbb{Z}}\oplus L_{mn}
\,,\,\,\,\,\,\,\,
L_{mn}=-\frac{d^2}{dt^2}+1+V_{mn}(t)\,,
\label{L}
\end{equation}
where $L_{mn}$ are the Schr\"{o}dinger operators with P\"{o}schel-Teller potentials
\begin{equation}
V_{mn}(t)=\left(k_{mn}^2+\frac{1}{2}\right){\rm sech}^2t+\left(m^2-\frac{1}{4}\right){\rm cosh}^2t\,,\,\,\,\,\,\,\,
k_{mn}:= -\frac{mb}{a}+\frac{\pi n}{a}.\label{eq: kmn}
\end{equation}
For details of this and the following remarks the reader can
consult \cite{Perry, Sjostrand, Guilope}, for example.
\footnote{
In particular, the proof of equation (\ref{L}) is given in Section 3 of \cite{Perry} - in equations (3.10)-(3.21) there.
}
The Schr\"{o}dinger equation
$
\Psi ''(x)+[E-V_{mn}(x)]\Psi(x)=0,
$
which is the same as the eigenvalue problem $L_{mn}\Psi=k^2\Psi$ for
$E=k^2-1$, has a known solution $\Psi^+(x)$ (in terms of the
hypergeometric function) with asymptotics
\begin{equation}
\Psi^+(x)\sim
\frac{e^{ikx}}{T_{mn}(k)}+
\frac{R_{mn}(k)}{T_{mn}(k)}e^{-ikx},
\end{equation}
for
{\it reflection} and {\it transmission} coefficients
$T_{mn}(k)$, $R_{mn}(k)$ respectively.
\begin{definition}
For $s$ defined by $k=i(1-s)$ one can form the scattering matrix
$
\left[{\frak S}_{mn}(s)\right]\stackrel{def}{=}
\left[R_{mn}(k)\right]
$
of $-\Delta_\Gamma$, whose entries are quotients of gamma
functions with "trivial poles" $s=1+j$,
$j=0,1,2,3,\mathellipsis,$ and non-trivial poles
\begin{equation}
s_{mnj}^{\pm}:= -2j-\left|m\right|\pm i\left|k_{mn}\right|
\label{poles}
\end{equation}
for $k_{mn}$ in (\ref{eq: kmn}), $j=0,1,2,3,\mathellipsis$.
The $s^{\pm}_{mnj}$ are the scattering resonances.
\end{definition}

\section{Truncated heat kernel and  the zeta function}

In this section we briefly relate the result of the heat kernel trace (integration over the fundamental domain $F=F_{(a,b)}$ along the diagonal) and a zeta function $Z_\Gamma(s)$ which has been discussed in \cite{Bytsenko1}
The calculation is carried out conveniently with spherical
coordinates: for $\rho\geq 0,\,\, 0\leq\theta\leq 2\pi,\,\, 0\leq\phi\leq\pi/2,\,\, x=\rho \sin{\phi} \cos{\theta},\,\, y=\rho \sin{\phi}\sin{\theta},\,\, z=\rho \cos{\phi}$.
\begin{notation}
For $p=(x,y,z)\in H^3$,
its image (i.e. its $\Gamma$-orbit) under the quotient map
$H^3\rightarrow\Gamma\backslash H^3= {\bf H}_\Gamma$ will be denoted by $\tilde{p}$. $d(p_{1},p_{2})$ will denote the hyperbolic
distance between two points
$p_{1},\,p_{2}=(x_{1},y_{1},z_{1}),\,(x_{2},y_{2},z_{2})$ in $H^3$:
\begin{equation}
\label{eq: coshdp1p2}\cosh{ d(p_{1},p_{2})} := 1+\frac{(x_{1}-x_{2})^2+(y_{1}-y_{2})^2+
(z_{1}-z_{2})^2}{2z_{1}z_{2}}\,.
\end{equation}
\end{notation}
The heat kernel $K^\Gamma_t$ ($t>0$) of ${\bf H}_\Gamma$ is obtained by averaging over $\Gamma$ the heat kernel $K_t$ of $H^3$:
\begin{equation}
K^\Gamma_t\left(\widetilde{p_{1}},\widetilde{p_{2}}\right) = \sum_{n\in\mathbb{Z}}
K_t\left(p_{1},\gamma^n p_{2}\right)
= \sum_{n\in\mathbb{Z}}\frac{e^{-t-d\left(p_{1},\gamma^n
p_{2}\right)^2/ 4t}d\left(p_{1},\gamma^n p_{2}\right)}{(4\pi
t)^\frac{3}{2}\sinh{d\left(p_{1},\gamma^n
p_{2}\right)}}\,,
\label{Kt}
\end{equation}
where we write $\gamma$ for $\gamma_{(a,b)}$.
\begin{definition}
For later purposes it is convenient to set
$
\ell \stackrel{def}{=} 2a=2\pi r_{+}/\sigma,\,\, \theta \stackrel{def}{=} 2b=2\pi\left|r_{-}\right|/\sigma\,.
$
\end{definition}

Let $K_t^{*\Gamma}(\widetilde{p_{1}},\widetilde{p_{2}})$ denote the {\it truncated} heat kernel of ${\bf H}_\Gamma$, defined by
restricting the sum over $\mathbb{Z}$ in (\ref{Kt}) to the non-zero integers $n$. The following result for the trace of
$K_t^{*\Gamma}(\widetilde{p_{1}},\widetilde{p_{2}})$ holds:
\begin{theorem} [{\it A. A. Bytsenko, M. E. X. Guimar\~aes and 
F. Williams} \cite{Bytsenko1}]
For the volume element $dV=\frac{\sigma^3}{z^3}dxdydz$, $t>0$,
and the theta-function
\begin{equation}
\Theta_{\Gamma}(t) :=
\frac{\ell}{8\sqrt{4\pi t}}
\sum_{n\in\mathbb{Z}-\{0\}}\frac{e^{-\left(t+\frac{n^2\ell^2}{4t}\right)}}{[\sinh^2\left(\frac{\ell n}{2}\right)+
\sin^2\left(\frac{\theta n}{2}\right)]}
= \frac{\ell}{4\sqrt{4\pi t}}\sum_{n=1}^\infty
\frac{e^{-\left(t+\frac{n^2\ell^2}{4t}\right)}}
{[\sinh^2\left(\frac{\ell n}{2}\right)+\sin^2\left(\frac{\theta
n}{2}\right)]},
\label{Th}
\end{equation}
one has that
\begin{equation}
{{\int\!\!\!\int\!\!\!\int}} K_{t}^{*\Gamma}(\widetilde{p},\widetilde{p})dV =2\sigma^3\Theta_{\Gamma}(t).
\label{int}
\end{equation}
\end{theorem}
The following zeta function has been attached to the BTZ black hole ${\bf H}_\Gamma$ (see for detail \cite{Perry,Williams}):
\begin{equation}
Z_\Gamma(s) :=\prod_{\stackrel{k_1,k_2\geq
0}{k_1,k_2\in\mathbb{Z}}}^\infty[1-(e^{i\theta})^{k_1}(e^{-i\theta})^{k_2}e^{-(k_1+k_2+s)\ell}]\,.
\label{zeta}
\end{equation}
$Z_\Gamma(s)$
is an entire function of $s$, whose zeros are precisely the following complex numbers
$
\zeta_{n,k_{1},k_{2}} = -\left(k_{1}+k_{2}\right)+i\left(k_{1}-
k_{2}\right)\theta/\ell+ 2\pi  in/\ell$, and whose logarithmic derivative for ${\rm Re} s> 0$ is given by
\begin{equation}
\frac{d}{ds}{\rm log} Z_\Gamma (s) =
\frac{\ell}{4}\sum_{n=1}^{\infty}\frac{e^{-n\ell(s-1)}}
{[\sinh^2\left(\frac{\ell n}{2}\right)
+\sin^2\left(\frac{\theta n}{2}\right)]}\,.
\label{Z2}
\end{equation}
\begin{remark}
It is remarkable fact that the set of scatering poles in (\ref{poles}) coincides with zeros of $\zeta_{n,k_{1},k_{2}}$,
as it can be verified. Thus encoded in $Z_{\Gamma}(s)$ is the spectrum of a BTZ black hole.
\end{remark}
$Z_\Gamma(s)$ is connected with the theta function $\Theta_\Gamma(t)$ in
(\ref{Th}), and hence with the
heat kernel $K_t^{*\Gamma}$ via the following
theorem:
\begin{theorem} [{\it A. A. Bytsenko, M. E. X. Guimar\~aes and 
F. Williams} \cite{Bytsenko1}]
For ${\rm Re}s>1$ one has
\begin{equation}
\int_0^\infty e^{-s(s-2)t}\Theta_\Gamma(t)dt=\frac{1}{2(s-1)}
\frac{d}{ds}{\rm log}Z_\Gamma(s)\,.
\end{equation}
\end{theorem}

\section{Quantum corrections}

We set up some more notation. For $\sigma_p$ the natural representation of $SO(2k-1)$ on space of $p-$forms $\Lambda^p {\Bbb C}^{2k-1}$ one has the corresponding Harish-Chandra-Plancherel density $\mu_{\sigma_p(r)}$ given, for a suitable normalization of 
Haar measure $dx$ on $G$. Simplifying calculations we will take into account the case of smooth functions {\rm (}$p=0${\rm )} or smooth vector bundle sections, and therefore the measure $\mu(r)\equiv \mu_{0}(r)$ corresponds to the trivial representation of $SO(n-1)$. Let $\chi_{\sigma} = {\rm trace}(\sigma)$ be the character of $\sigma$. Since $\sigma_0$ is the trivial representation one has $\chi_{\sigma_0}=1$.  
It follows that the scalar determinant on ${\bf H}_\Gamma$
has the form
\begin{eqnarray}
{\rm log}\,{\rm det}\,\triangle_\Gamma & = & -\int_o^{\infty}t^{-1}
{\rm Tr} K^{*\Gamma}_t\left(\widetilde{p_{1}},
\widetilde{p_{2}}\right)dt
= -2\sigma^3\int_0^{\infty}t^{-1}\Theta_\Gamma(t)dt
\nonumber \\
& = & -2\sigma^3\int_0^{\infty}
\sum_{n=1}^{\infty}\frac{\ell}{4t\sqrt{4\pi t}}
\frac{e^{-(t+\frac{n^2\ell^2}{4t})}dt}
{[\sinh^2\left(\frac{\ell n}{2}\right)
+\sin^2\left(\frac{\theta n}{2}\right)]}\,.
\label{integral}
\end{eqnarray}
The calculation of the integral in (\ref{integral}) relies on the following reprezentation for the Bessel function $K_{\nu}(s)$,
\begin{equation}
\int_{0}^{\infty}x^{\nu-1}e^{-\frac{\alpha}{x}-\beta x}dx
= 2\left(\frac{\alpha}{\beta}\right)^{\nu/2}\!K_{\nu}(2\sqrt{\alpha \beta})\,,
\,\,\,\,\,\,\, {\rm Re}\, \alpha >0\,,\,\,\,{\rm Re}\, \beta >0\,.
\end{equation}
Taking into account that $K_{\pm 1/2}(s)= \sqrt{\pi/2s}\exp(-s)$ we finally get
\begin{equation}
{\rm log}\,{\rm det}\,\triangle_\Gamma = 
-\frac{\sigma^3}{2}\sum_{n=1}^{\infty}
\frac{e^{-n\ell}}
{n[\sinh^2\left(\frac{\ell n}{2}\right)
+\sin^2\left(\frac{\theta n}{2}\right)]}
= 2\sigma^3{\rm log}Z_\Gamma(2)\,.
\end{equation}

\subsection{The tensor kernel and spectral functions on p-forms}
Quantum corrections coming from small fluctuations around 
metric extremum and from gauge-fixing can be computed by taking advantage of the of the connection between three-dimensional gravity and Chern-Simons theory \cite{Witten,Carlip1}.
One-loop corrections can be given by the square root of the
Ray-Singer torsion \cite{Ray} (or equivalently the
Reidemeister-Franz torsion) of the manifold ${\bf H}_\Gamma$.
Let $\xi$ be an irreducible representation of $K$ on a
complex vector space $V_\xi$, and form the induced homogeneous
vector bundle $G\times_K V_\xi$ (the fiber product of
$G$ with $V_\xi$ over $K$). Restricting the $G$ action to $\Gamma$ we obtain the quotient bundle
\begin{equation}
E_\xi =\Gamma\backslash(G\times_KV_\xi)\longrightarrow
{\bf H}_\Gamma\,.
\label{bundle}
\end{equation}
The natural Riemannian structure on ${\bf H}_\Gamma$ induced by the Killing form $(,)$ of $G$ gives rise to a connection Laplacian $\triangle_\Gamma$ on $E_\xi$.
Let us briefly recall the definition of the Ray-Singer torsion.
Suppose ${\mathbb E}= (E, \nabla)$ is a real or complex vector
bundle $E$ equipped with a flat connection $\nabla$. Let $g$ be a Riemannian metric on a manifold $M$ and let $\mu$ be a Hermitian metric on $E$.
\begin{definition}
The Ray-Singer torsion $T_{an}(\nabla, g, \mu)$ is given by
\begin{equation}
{\rm log}T_{an}(\nabla, g, \mu)\stackrel{def}{=}
\frac{1}{2}\sum_p(-1)^{p+1}p\,{\rm log}[{\rm det}\triangle^{(p)}]_{reg}\,.
\label{torsion}
\end{equation}
Here $\triangle^{(p)}$ denotes the Laplacian in degree 
$p$ of the
elliptic complex $(\Omega(X; E), d_\nabla)$ when equipped with the scalar product
induced from the Riemannian metric $g$ and the Hermitian metric $\mu$, and
$[{\rm det}\triangle^{(p)}]_{reg}$ denotes its zeta regularized determinant.
\end{definition}
The bundle (\ref{bundle}) arises in the theory of geometric structures, and admits a natural flat connection with a holonomy group isomorphic to $\Gamma$. 
\footnote{
It has been shown \cite{Witten} that three-dimensional gravity
can be rewritten as a Chern-Simons theory for the gauge group $G$. In addition, extremum of the action determines a flat connection, with a corresponding bundle given by (\ref{bundle}). For the Chern-Simons invariant of an irreducible flat connection on real hyperbolic three-manifold see \cite{Bonora}.} 
If there are no ghost zero-modes, that is, if the de Rham cohomology $H^0({\bf H}_\Gamma; E_\omega) = 0$ on forms $\omega$ of ${\bf H}_\Gamma$ with values in the flat bundle $E_\omega$, then the Ray-Singer torsion is a topological invariant (is not dependent on the choice of auxiliary metric). When zero modes are present, they must be included in Eq. (\ref{torsion}).
\begin{remark}
A version of the trace formula for the heat kernel on $p-$forms, developed in {\rm \cite{Fried}} {\rm (}see also 
{\rm \cite{Bytsenko1}}{\rm )}, leads to the identity $I_\Gamma^{(p)}$ and geodesic ${\frak G}_\Gamma^{(p)}$ terms. This decomposition reduces to divergences of identity terms, since they proportional to ${\rm Vol}({\bf H}_\Gamma)$, and hence to divergent part of the effective action. Note that identity terms describe the renormalizsation of the cosmological constant at one-loop, and can be canceled by local counterterms. 
\end{remark}

\subsection{Quantum contribution to the partition functions}
\begin{remark}
The effective action for the scalar field on three-dimensional BTZ black hole instanton can be calculated as follows. The non-divergent part of the effective action is given by
{\rm \cite{Mann}}:
\begin{equation}
W_{(\rm non-divergent)}^{\,\,\rm scalar} = -\sum_{n=1}^{\infty}
\frac{e^{-\sqrt{1-\xi}n\ell}}
{4n[\sinh^2\left(\frac{\ell n}{2}\right)
+\sin^2\left(\frac{\theta n}{2}\right)]}\,,
\end{equation}
where the constant $(1-\xi)$ appearing in the heat equation formula. In the conformal invariant case of non-minimal coupling we have $\xi = 3/4$. Using the Selberg-like zeta function $Z_\Gamma(s)$
we get
\begin{equation}
W_{(\rm non-divergent)}^{\,\,\rm scalar} = {\rm log}\,Z_\Gamma(1+\sqrt{1-\xi})\,.
\end{equation}
The divergent part of the effective action could be supressed by introducing the local counterterms.
\end{remark}

Corrections to the spectrum of three-dimensional gravity and the
one-loop contribution to the partition function {\rm (}the contribution
of states of left- and right-moving modes{\rm)} of the conformal field
theory has been evaluated in {\rm \cite{Maloney,Giombi}}. The later
contribution has the form:
\begin{equation}
{\bf Z}(\tau) = 
|q \overline{q} |^{-k}
\prod_{n=2}^{\infty}|1-q^n|^{-2}\,,
\label{Z1}
\end{equation}
where $24k= c_L=c_R= c$, and $c$ is the central charge of a conformal field theory. In {\rm (\ref{Z1})} $q=\exp[2\pi i\tau]=\exp[2\pi(-{\rm Im}\tau +i{\rm Re}\tau)]$ such that 
$|\overline{q} q|^{-k}=\exp[4\pi k {\rm Im}\tau]$ corresponds to
the classical prefactor of the partition function.
To make correspondence between models one can choose the pair $(a,b)$ in our definition {\rm (\ref{g1})}, {\rm (\ref{G1})} as $(a=-\pi{\rm Im}\tau, b=\pi{\rm Re}\tau)$. Then since $k=(16\sigma G)^{-1}$
one gets
\begin{equation}
-{\rm log}{\bf Z}_{\rm cl}(\tau) = 
k{\rm log}|q \overline{q}| = 4\pi k\sigma r_+=
\frac{4\pi^2r_+}{16\pi G}\,.
\label{class}
\end{equation}
The result {\rm (\ref{class})} is the classical part of the contribution
{\rm (}see for example {\rm \cite{Bytsenko}}{\rm )}.
Eq. {\rm (\ref{Z1})} is one-loop exact as has been claimed in {\rm \cite{Maloney}}.

{\bf Modular invariance}.
For three-dimensinal gravity in locally Anti-de Sitter space-times the one-loop partition function has been calculate in \cite{Giombi}, and the result is (Eqs. (4.27) and (4.28) of \cite{Giombi}): 
\begin{equation}
{\bf Z}_{\rm gravity}^{\rm 1-loop}(\tau) = 
\prod_{m =2}^{\infty}|1-q^m|^{-2}\,.
\label{Z0}
\end{equation}
If we let $\ell = 2\pi {\rm Im}\,\tau,\, \theta= 
2\pi{\rm Re}\,\tau$, then
\begin{equation}
\sinh^2\left(\frac{\ell n}{2}\right)
+\sin^2\left(\frac{\theta n}{2}\right) = |\sin(n\pi \tau)|^2 =
\frac{|1-q^n|^2}{4|q|^{n}}\,.
\label{q}
\end{equation}
Using Eq. (\ref{q}) we get
\begin{eqnarray}
{\rm log}\prod_{m=2}^{\infty}|1-q^m|^{-2} & = & -\sum_{m=2}^{\infty} 
{\rm log}|1- q^m|^2 = \sum_{n=1}^{\infty}
\frac{q^{2n} + \overline{q}^{2n} - |q|^{2n}(q^n+ \overline{q}^n)}{n|1-q^n|^2} 
\nonumber \\
& = &
\sum_{n=1}^{\infty}\frac{e^{-2n\pi{\rm Im}\,\tau}\cos
(4n\pi {\rm Re}\,\tau) - e^{-4n\pi{\rm Im}\,\tau}\cos
(2n\pi {\rm Re}\,\tau)}
{2n[\sinh^2\left(n\pi {\rm Im}\tau\right)
+\sin^2\left(n\pi{\rm Re}\tau\right)]}\,.
\label{modular}
\end{eqnarray}
One can summarize the preceding formulas by geting the following result
\begin{eqnarray}
{\rm log}{\bf Z}_{\rm gravity}^{\rm 1-loop}(\tau) & = & 
\sum_{n=1}^{\infty}\frac{e^{-2n\pi{\rm Im}\,\tau}\cos
(4n\pi {\rm Re}\,\tau) - e^{-4n\pi{\rm Im}\,\tau}\cos
(2n\pi {\rm Re}\,\tau)}
{2n[\sinh^2\left(n\pi {\rm Im}\tau\right)
+\sin^2\left(n\pi{\rm Re}\tau\right)]}
\nonumber \\
& = &
{\rm log}\left[
\frac{Z_\Gamma(3+ i\frac{{\rm Re}\tau}{{\rm Im}\tau})
Z_\Gamma(3 - i\frac{{\rm Re}\tau}{{\rm Im}\tau})}
{Z_\Gamma(1+ i\frac{{\rm Re}\tau}{{\rm Im}\tau})
Z_\Gamma(1 - i\frac{{\rm Re}\tau}{{\rm Im}\tau})}
\right].
\label{final0}
\end{eqnarray}
Note that in \cite{Giombi} the one-loop partition function of three-dimensional gravity in hyperbolic spaces has been considered and stressed that such a geometry is also geometry of the Euclidean BTZ black hole. Thus if we let the modular transformation $\tau = \frac{1}{2\pi}(\theta + i\ell) \rightarrow -1/\tau$, i.e.
\begin{equation}
{\rm Re}\,\tau \longrightarrow - \frac{{\rm Re}\,\tau}
{|\tau|^{2}},\,\,\,
{\rm Im}\,\tau \longrightarrow  
\frac{{\rm Im}\,\tau}{|\tau|^{2}},\,\,\, 
\frac{{\rm Re}\,\tau}{{\rm Im}\,\tau} \longrightarrow - 
\frac{{\rm Re}\, \tau}{{\rm Im}\,\tau}\,,
\label{transform}
\end{equation}
then the computation of the one-loop partition function of the three dimensional gravity gives the one loop correction to black holes. It is easy seen that the last formula in (\ref{final0}) is invariant under the transformation $\tau \rightarrow 1+\tau$. Therefore (we let $\sigma = 1$)
\begin{equation}
{\bf Z}_{\rm gravity}^{\rm 1-loop}(\tau)
={\bf Z}_{(\rm spinning\,\,\, BTZ)}^{\rm 1-loop}(\ell, \theta)\,.
\end{equation}
Finally, simplifying calculations one can consider a non-spinning black hole (J=0); in this case $\theta = 0\, 
({\rm Re}\tau = 0)$ and
\begin{equation}
{\bf Z}_{(\rm non-spinning\,\,\, BTZ)}^{\rm 1-loop}=
\left[\frac{Z_\Gamma^{\theta=0}(3)}
{Z_\Gamma^{\theta=0}(1)}\right]^2\,.
\end{equation}

\section{Concluding remarks}

In this paper we obtained the quantum correction to the BTZ black hole making use of three-dimensional gravity representation \cite{Maloney, Giombi}.
We have studied the quantum correction by expressing the determinants in terms of the appropriate heat kernels, which were evaluated using a method of images. This correction was expressed as a special value of the logarithms of the zeta functions. 
Earlie the first quantum correction for non-spinning black hole ($J=0$) has been obtained in \cite{Bytsenko}, but it has error in the computation. In this paper we revised the correction using 
a special relation between the spectrum and the truncated heat kernel of the Euclidean BTZ spinning black hole with the Patterson-Selberg zeta function. It gives us possibility to provide the correct result for the quantum correction.

\subsection*{Acknowledgments}

A. A. Bytsenko and M. E. X. Guimar\~aes would like to thank the
Conselho Nacional de Desenvolvimento Cient\'ifico e Tecnol\'ogico
(CNPq) for  support.


\begin{thebibliography}{99}

\bibitem{Maloney}
A. Maloney and E. Witten, {\it Quantum Gravity Partition
Function In Three Dimensions}, arXiv:0712.0155.

\bibitem{Giombi}
S. Giombi, A. Maloney and X. Yin, {\it One-loop Partition Functions of 3D Gravity}, arXiv:hep-th/0804.1773.

\bibitem{Banados}
M. Ba\~nados, C. Teitelboim and J. Zanelli, {\it Black hole in three-dimensional spacetime}, Phys. Rev. Letters {\bf 69} (1992)
1849-1851.

\bibitem{Perry}
P. Perry and F. Williams, {\it Selberg zeta function and trace formula for the BTZ black hole}, Internat. J. of Pure and Applied Math. {\bf 9} (2003) 1-21.

\bibitem{Perry1}
P. Perry, {\it A Poisson summation formula and lower bounds for resonances in hyperbolic manifolds}, Int. Math. Res. Notes
{\bf 34} (2003) 1837-1851.

\bibitem{Patterson1}
S. J. Patterson, {\it The Selberg zeta-function of a Kleinian group}, In Number Theory, Trace Formulas, and Discrete groups: Symposium in Honor of Atle Selberg, Oslo, Norway, July 14-21, 1987. New York, Academic press, 1989.

\bibitem{Patterson2}
S. J. Patterson and P. A. Perry, {\it The deivisor of the Selberg zeta function for Kleinian groups, with an appendix by Charles Epstein}, Duke Math. J. {\bf 106} (2001) 321-390. 

\bibitem{Bytsenko}
A. A. Bytsenko, L. Vanzo, S. Zerbini, {\it Quantum Correction to the Entropy of the (2+1)-Dimensional Black Hole},
Phys. Rev. D {\bf 57} (1998) 4917-4924; [arXiv:gr-qc/9710106].

\bibitem{Bytsenko2}
A. A. Bytsenko, L. Vanzo, S. Zerbini, {\it Ray-Singer Torsion for a Hyperbolic 3-Manifold and Asymptotics of Chern-Simons-Witten Invariant}, Nucl. Phys. B {\bf 505} (1997) 641-659;
[arXiv:hep-th/9704035].

\bibitem{Bytsenko1}
A. A. Bytsenko, M. E. X. Guimaraes and F. L. Williams,
{\it Spectral Functions for BTZ Black Hole Geometry},
Lett. Math. Phys. {\bf 79} (2007) 203-211; [arXiv:hep-th/0609102].

\bibitem{Sjostrand}
J. Sj\"{o}strand and M. Zworski, {\it Lower bounds on the number of scattering poles}, Comm. Partial Diff. Equations {\bf 18} (1993) 847-857.

\bibitem{Guilope}
L. Guillop\'{e} and M. Zworski, {\it Upper bounds on the number of resonances for non-compact Riemann surfaces}, J. of Funct. Analysis {\bf 129} (1995) 364-389.

\bibitem{Williams}
F. Williams, {\it A zeta function for the BTZ black hole},  Internat. J. Modern Physics A {\bf 18} (2003) 2205-2209.

\bibitem{Witten}
E. Witten, {\it 2+1 Dimensional gravity as an exactly soluble system}, Nucl. Phys. B {\bf 311} (1988/89) 46-78.

\bibitem{Carlip1}
S. Carlip, {\it The Sum over Topologies in Three-Dimensional Euclidean Quantum Gravity}, Class. Quant. Grav. {\bf 10} (1993) 207-218; [arXiv:hep-th/9206103].

\bibitem{Ray}
D. B. Ray and I. M. Singer, {\it Analytic Torsion For Complex Manifolds}, Annals Math. {\bf 98} (1973) 154-177.

\bibitem{Bonora}
L. Bonora and A. A. Bytsenko, {\it Fluxes, Brane Charges and Chern Morphisms of Hyperbolic Geometry}, Class. Quant. Grav.
{\bf 23} (2006) 3895-3916; [arXiv:hep-th/0602162].

\bibitem{Fried}
D. Fried, {\it Analytic torsion and closed geodesics on hyperbolic manifolds}, Invent. Math. {\bf 84} (1986) 523-540.

\bibitem{Mann}
R. B. Mann and S. N. Solodukhin, {\it Quantum scalar field on three-dimensional (BTZ) black hole instanton: heat kernel, effective action and thermodynamics}, Phys. Rev. D {\bf 55}
(1997) 3622-3632; [arXiv:hep-th/9609085].





\end{thebibliography}
\end{document}